\documentclass[twocolumn,showpacs,preprintnumbers,amsmath,amssymb]{revtex4}
\usepackage{graphicx}
\usepackage{color}
\usepackage{dcolumn}
\usepackage{bm}

\begin{document}
\title{Group-wave resonances in nonlinear dispersive media: The case of gravity water waves}
\author{A.V. Slunyaev$^{1,2}$}

\affiliation{$^1$Institute of Applied Physics, 46 Ulyanova Street, N. Novgorod 603950, Russia \\
$^2$Nizhny Novgorod State Technical University, 24 Minina Street, N. Novgorod 603950, Russia}

\date{\today}

\begin{abstract}
The dynamics of coherent nonlinear wave groups is shown to be drastically different from the classical scenario of weakly nonlinear wave interactions. The coherent groups generate non-resonant (bound) waves which can be synchronized with other linear waves. By virtue of the revealed mechanism the groups may emit waves with similar or different lengths, which propagate in the same or opposite direction.

\end{abstract}

\maketitle
\emph{Introduction.---}
Waves are resonant when the combination of their Fourier phases exhibits no or slow variation. This holds when
\begin{align} \label{ResonanceCondition}
\sum_j{\pm k_j} = 0 \quad \text{and} \quad \sum_j{\pm \omega_j} = 0, \qquad j = 1, 2, ... ,
\end{align}
where $j$ counts the interacting waves,
$k_j$ are wavenumbers, and $\omega_j$ are the corresponding frequencies. One spatial dimension is considered for the sake of simplicity.
%
Waves may efficiently exchange energy when the resonance conditions (\ref{ResonanceCondition}) are satisfied exactly or approximately. The role of near-resonant interactions is not less important as of the exact resonances \cite{StiassnieShemer2005}.
Meanwhile the resonance conditions in nonlinear systems are still formulated for linear frequencies (e.g., \cite{Kartashova2010}) having in mind the weakly nonlinear limit.

In application to water waves, and in particular oceanic waves,
the vast modern machinery of the wind-generated wave modeling is based on the weakly nonlinear kinetic equations which account for the interactions in resonant quadruplets  \cite{WISE2007}. The predominance of four-wave processes originates from the linear dispersion relation, which for gravity waves over infinitively deep water reads
%
$\omega^2 = g |k|$,
%
where $g$ is the acceleration due to gravity.
%
%
Some profound conclusions follow from the theoretical analysis of the properties of the four-wave coupling coefficient.
So, when the surface waves are collinear, the nolinear coefficient for non-trivial resonant 4-wave interactions cancels out, initially unidirectional waves cannot excite waves in the opposite direction \cite{DyachenkoZakharov1994,Zakharov1999}.

Waves may receive energy from moving objects due to the Cherenkov resonance, when the velocity of the object, $V$, is equal to the phase velocity, $\omega / k$, of the resonant wave.
%
%
Localized wave groups (and, in particular, quasi-solitons) may interact with waves according to this scenario \cite{ZakharovKuznetsov1998}. Then the resonance condition has the form
\begin{align} \label{CherenkovResonanceGroup}
V = (\omega - \omega_0) / (k-k_0) ,
\end{align}
where $k_0$ and $\omega_0$ are the dominant wavenumber and frequency of the group, and $V$ is the group velocity. Its graphical interpretation will be given below. It may be also constructive to rewrite condition (\ref{CherenkovResonanceGroup}) in the form of equality of the two Doppler-shifted frequencies, $\omega - kV = \omega_0 - k_0 V$.

The potential role of the modulational instability in the extreme dynamics of oceanic waves has been discovered recently \cite{Onoratoetal2009}. It leads to occurrence of soliton-like coherent wave groups, which essentially alter the wave statistics.
Meanwhile soliton-like intense groups of water waves have been first observed in numerical simulations, and then reproduced in laboratory conditions \cite{DyachenkoZakharov2008}.
The issue of stability of these groups has been a mere blank up till now.

In this letter we present the evidences of essentially non-classical processes of the energy exchange between the strongly modulated coherent wave groups and quasi-linear waves, which exist in addition to the classic nonlinear wave resonances.  Besides we show that the picture of distribution of energy in the Fourier domain, which, in particular, determines the wave spectra which can be measured instrumentally, is drastically different from that which is conventionally assumed. 

\begin{figure}
\centerline{\includegraphics[width=9.0cm]{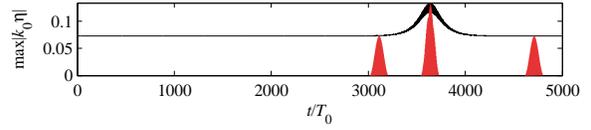}}
\caption{The maximum displacement $\eta$ as the function of time (black line) and three selections by the windows $M$ (the red areas), the case $\epsilon = 0.07$, $L/\lambda=6$.}
\label{fig:Maxima}
\end{figure}

\emph{Setup for the fully nonlinear simulation of modulated water waves.---}
We consider a very general setting for collinear gravity waves on the water surface, assuming a potential flow of ideal incompressible fluid described by the velocity potential $\phi(x,z,t)$ and limited from above by the free boundary $\eta(x,t)$,
\begin{align} \label{EulerEq}
\phi_t + \frac{1}{2}\left(\phi_x^2 + \phi_z^2 \right) + g \eta = 0  \quad \text{at} \quad z = \eta, \\
\eta_t + \eta_x \phi_x = \phi_z  \quad \text{at} \quad z = \eta, \nonumber  \\
\phi_{xx} + \phi_{zz} = 0, \quad - \infty < z \leq \eta, \nonumber \\
\phi_z \rightarrow 0 \quad \text{when} \quad z \rightarrow - \infty. \nonumber
\end{align}
When the classic Euler equations (\ref{EulerEq}) are written in conformal variables, they may be solved numerically by a pseudo-spectral method with a high accuracy having no assumptions on the degree of the wave nonlinearity or the spectral width \cite{Zakharovetal2002}. The evolution in time is calculated by means of the pseudo-spectral numerical algorithm with the help of the  4-order Runge-Kutta method. The robustness of the simulations with respect to finer discrete grids and other integration algorithms was verified.

The initial condition for the numerical simulation is a numerically exact stationary uniform periodic wave (the Stokes wave) with the height $H$ and the period $\lambda$ in the periodic computational domain of the length $L$. The important parameters of the numerical experiments are the wave steepness, $\epsilon \equiv k_0 H/2 = O(10^{-1})$, where $k_0 = 2\pi\lambda$, and the relative length of modulation, $L/\lambda$. The Stokes wave is known to be unstable with respect to the modulational instability.
The parameters of the initial condition are selected in such a way that the train is modulationally unstable, and only one mode of the instability may develop.
Seeding perturbations slightly exceeding the level of numerical noise are introduced at the unstable wavenumber of the initial condition  with the purpose to initiate the modulational growth.

The analysis of the space-time Fourier amplitudes $S(k,\omega)$ will be the main tool for examining the wave dynamics. The double Fourier transforms of the surface displacement $\hat{F} \{ \eta \}$ are calculated for the time subdomains of the simulated data with the use of the window function of time $M(t)$,
\begin{align} \label{Fourier}
S(k, \omega) = | \hat{F} \{ \eta(x,\tau) M(t-\tau)\} |, \\
M(t) = \frac{1}{2}- \frac{1}{2} \cos{ \frac{2\pi t}{W}}, \quad t \in [0,W) .
\end{align}
The Hanning function $M$  selects the time slice of the length $W$ and smoothly swamps the data at the window edges and hence improves the efficiency of the Fourier representation in time (see Fig.~\ref{fig:Maxima}). As the analyzed field $\eta$ is real-valued, the Fourier transform possess the symmetry property  $S(-k, -\omega) = S(k, \omega)$; therefore only the half-plane $\omega \geq 0 $ will be discussed in what follows.

\begin{figure}
\centerline{\includegraphics[width=8cm]{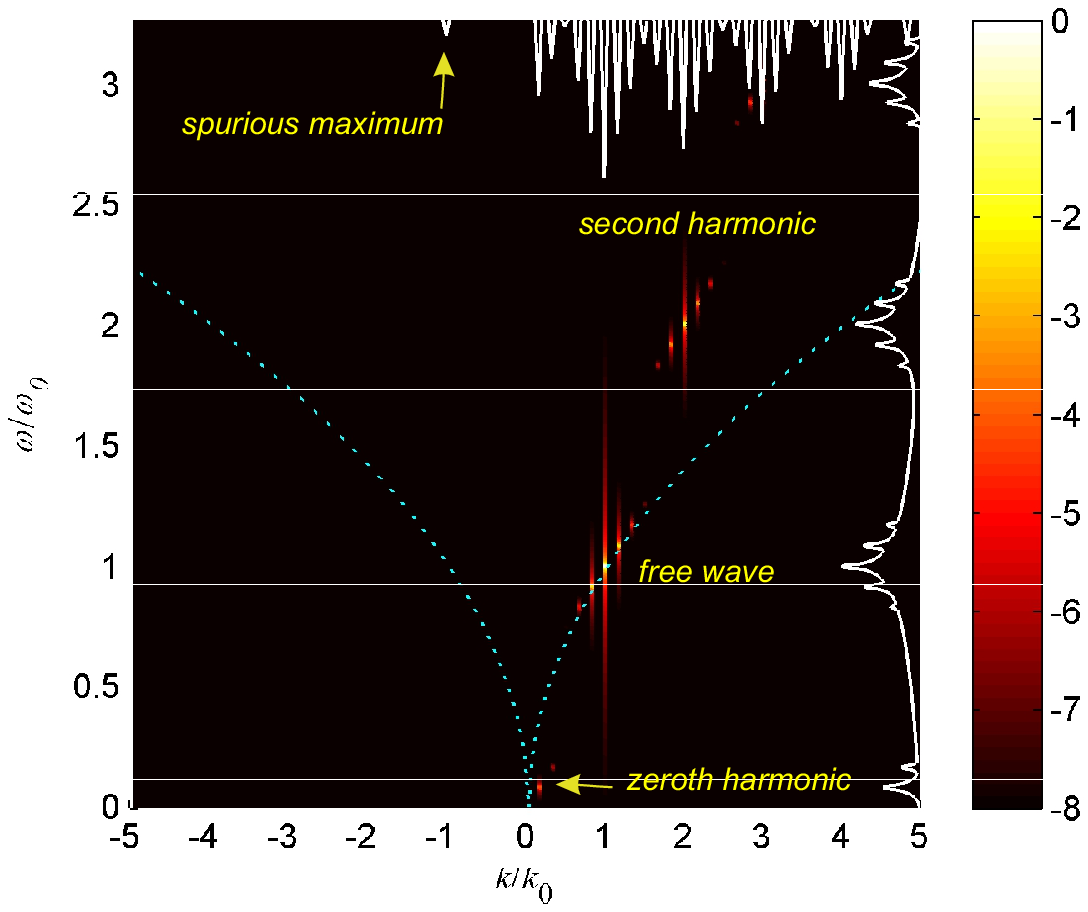}(a)}
\centerline{\includegraphics[width=8cm]{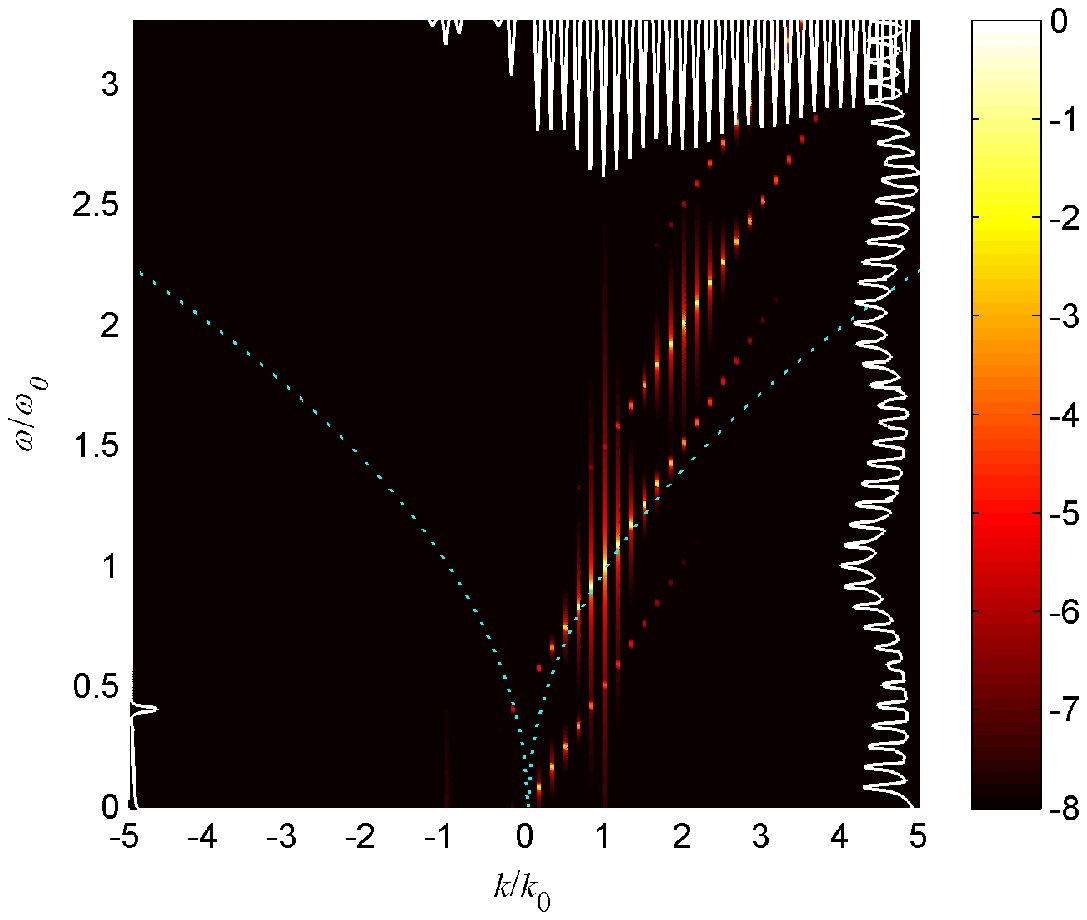}(b)}
\centerline{\includegraphics[width=8cm]{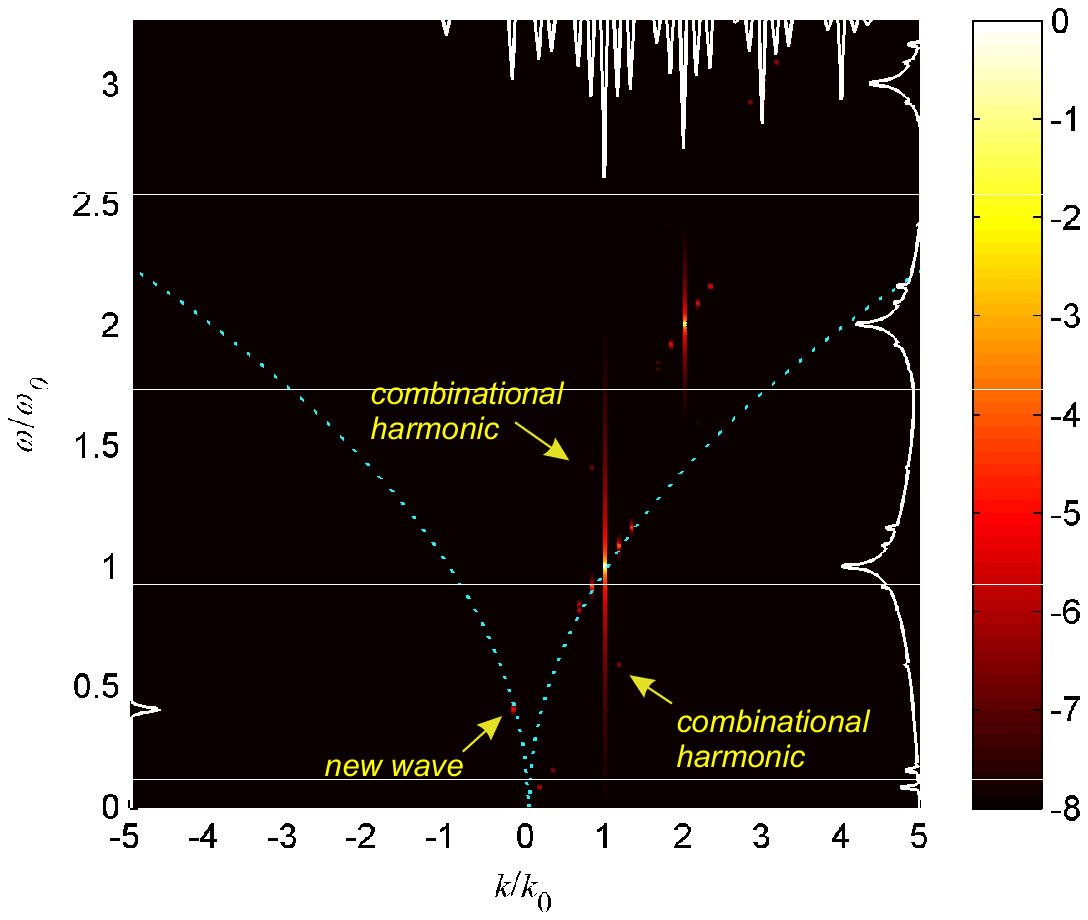}(c)}
\caption{The space-time Fourier transform amplitudes $\log_{10}{S(k,\omega)}$ are shown by color well before the first focusing ($t \approx 3100 T_0$) (a), at the moment of focusing ($t \approx 3600 T_0$) and when demodulated ($t \approx 4700 T_0$) (c), the case $\epsilon = 0.07$, $L/\lambda=6$.
The white curves show the Fourier transform amplitudes $\log_{10}{S_{\omega+}}$ (the right side), $\log_{10}{S_{\omega-}}$ (the left side) and $\log_{10}{S_k}$ (the upper side). The cyan dotted lines plot the dispersion curves. See clip \cite{MovieCase007}.}
\label{fig:XTSp007}
\end{figure}

The modulational instability is usually considered in either space or time domain, while Fig.~\ref{fig:XTSp007} reports on the joint space-time behaviour. The Fourier amplitudes $S(k,\omega)$ normalized by the maximum value are shown with color in the logarithmic scale; the relative values less than $10^{-8}$ are disregarded. By virtue of the integration of $S^2$ along the $k$ or $\omega$ axis the spatial Fourier transform, $S_k$, and the frequency Fourier transforms for the waves with positive wavenumbers, $S_{\omega+}$, and negative wavenumbers,  $S_{\omega-}$, are calculated

\emph{Interpretation of the $k-\omega$ plots.---}
As expected, the modulation instability develops leading to the occurrence of waves with about doubled amplitudes (Fig.~\ref{fig:Maxima}). The instability saturates at some stage, and a quasi-recurrent dynamics is observed eventually. The stage at about $500$ wave periods prior to the focusing (see the leftmost selected area in Fig.~\ref{fig:Maxima}) is shown in Fig.~\ref{fig:XTSp007}a.
The main peak of $S$ is located at the wavenumber $k_0$ and at the frequency $\omega_p$, which due to the nonlinearity is slightly larger than $\omega_0 \equiv \sqrt{g k_0}$ prescribed by the linear theory.
The cyan dotted lines plot the dispersive curves $\omega = \sqrt{g|k|}$ for waves propagating rightwards ($k > 0$) and leftwards ($k < 0$).

%


The locus of free (true) waves in the panel $(k, \omega)$ must correspond to the location of the dispersive curve (with account of the nonlinear shift). Other non-zero values of $S$ correspond to the bound (or phase-locked) waves which are present in the Fourier domain due to the fact that the progressive water waves are not sinusoidal. The bound waves may be eliminated after an appropriate canonical transformation within the perturbative Hamiltonian approach \cite{Zakharov1999}; they do not propagate without the 'parent' free waves.
The $n$-th order harmonics of the carrier wave are represented by peaks located at $(nk_0, n\omega_p)$, $n=2, 3, ...$. The second harmonic may be easily seen in Fig.~\ref{fig:XTSp007}a; the third harmonic is almost invisible in the $k-\omega$ plane, but is clearly observed in $S_k$ and $S_{\omega+}$.

A small-amplitude peak may be noticed in $S_k$ at $k=-k_0$; the corresponding peak in $S_{\omega-}$ is located very close to $\omega = 0$, and cannot be distunguished in Fig.~\ref{fig:XTSp007}a. This Fourier component originates from the maximum of $S$ at the conjugated point $(-k_0, -\omega_p)$, which has slowly decaying tails along the $\omega$ axis (it is better seen in Fig.~\ref{fig:XTSp0146}b below). This is an artefact of the processing; no opposite waves exist in Fig.~\ref{fig:XTSp007}a, at least with the relative amplitudes exceeding $10^{-8}$.

The modulation is clearly seen in functions $S_k$ and $S_{\omega+}$. In the panel $(k, \omega)$ the modulation of the free wave component is a line of peaks passing through $(k_0, \omega_p)$ close to the tangent to the dispersive curve. Only discrete wavenumbers $j k_0 \lambda /L$, where $j$ is integer, are allowed due to the imposed periodic boundary condition in space. The departure from the dispersive curve reveals the nonlinear character of the modulation and distinguishes it from linear dispersive trains.
The modulational instability may be interpreted as a result of a resonant interaction between two carrier waves $(k_0,\omega_p)$ and the satellites $(k_0 + \Delta k, \omega_p+\Delta \omega)$ and $(k_0 - \Delta k, \omega_p-\Delta \omega)$.
The alignment of the modulation peaks in the $k-\omega$ plane gives the evidence of the emergence of a coherent wave train with celerity $\Delta \omega/\Delta k$ close to the linear group wave velocity; this train corresponds to the fully nonlinear counterpart of the celebrated breather solution for the envelope equations \cite{SlunyaevShrira2013}.

The second and third-order bound waves in Fig.~\ref{fig:XTSp007}a qualitatively reproduce the modulation of the free waves centered at the peaks $(nk_0,n\omega_p)$.
The large-scale induced displacement  is represented in Fig.~\ref{fig:XTSp007}a by two peaks of the surface $S(k, \omega)$  near the origin; they are also seen in functions $S_{\omega+}$ and $S_k$.

The stage of the maximal wave amplification (see the second selection in Fig.~\ref{fig:Maxima}) is displayed in Fig.~\ref{fig:XTSp007}b.
The modulation is much stronger developed than compared with Fig.~\ref{fig:XTSp007}a: the number of observable spots in the bound waves lobes is greater, their amplitudes are larger. However, they still form straight lines in the $k-\omega$ plane, thus the modulated group propagates coherently as a whole. Different octaves overlap in the functions $S_k$ and $S_{\omega+}$, but still may be well separated in $S(k,\omega)$.

\emph{Generation of the opposite waves.---}
Note that in Fig.~\ref{fig:XTSp007}b there is one energetic peak with the negative wavenumber $k/k_0=-1/6$, which is readily seen in both, $S_k$ and $S_{\omega-}$. It belongs to the tail of the modulation of the free wave (the associated straight line passes $(k_0,\omega_p)$) and  is located close to the dispersive curve for opposite waves. This geometrical interpretation of the process may be straightforwardly related with the condition (\ref{CherenkovResonanceGroup}). The right hand side of (\ref{CherenkovResonanceGroup}) corresponds to the choice of shifted references $k^\prime = k - k_0$, $\omega^\prime = \omega - \omega_p$; then (\ref{CherenkovResonanceGroup}) represents the traditional Cherenkov resonance condition $V = \omega^\prime/k^\prime$, with $V$ being the velocity of the group, which corresponds to the incline of the series of energetic spots in the $(k, \omega)$ (or, equivalently,  $(k^\prime, \omega^{\prime})$) domain.
Later on during the partial demodulation (the third selected area in Fig.~\ref{fig:Maxima}) the excited opposite wave remains, see Fig.~\ref{fig:XTSp007}c, its amplitude is $O(\epsilon^5)$ ($\epsilon = O(10^{-1}$) of the amplitude of the carrier wave.

The case shown in Fig.~\ref{fig:XTSp007} corresponds to the classic Cherenkov radiation by a strongly localized steep wave group, which emerged due to the modulational instability.
The emitted wave propagates opposite to the dominant wave train, thus the originally unidirectional character of the wave dynamics gets broken.

\begin{figure}
\centerline{\includegraphics[width=8cm]{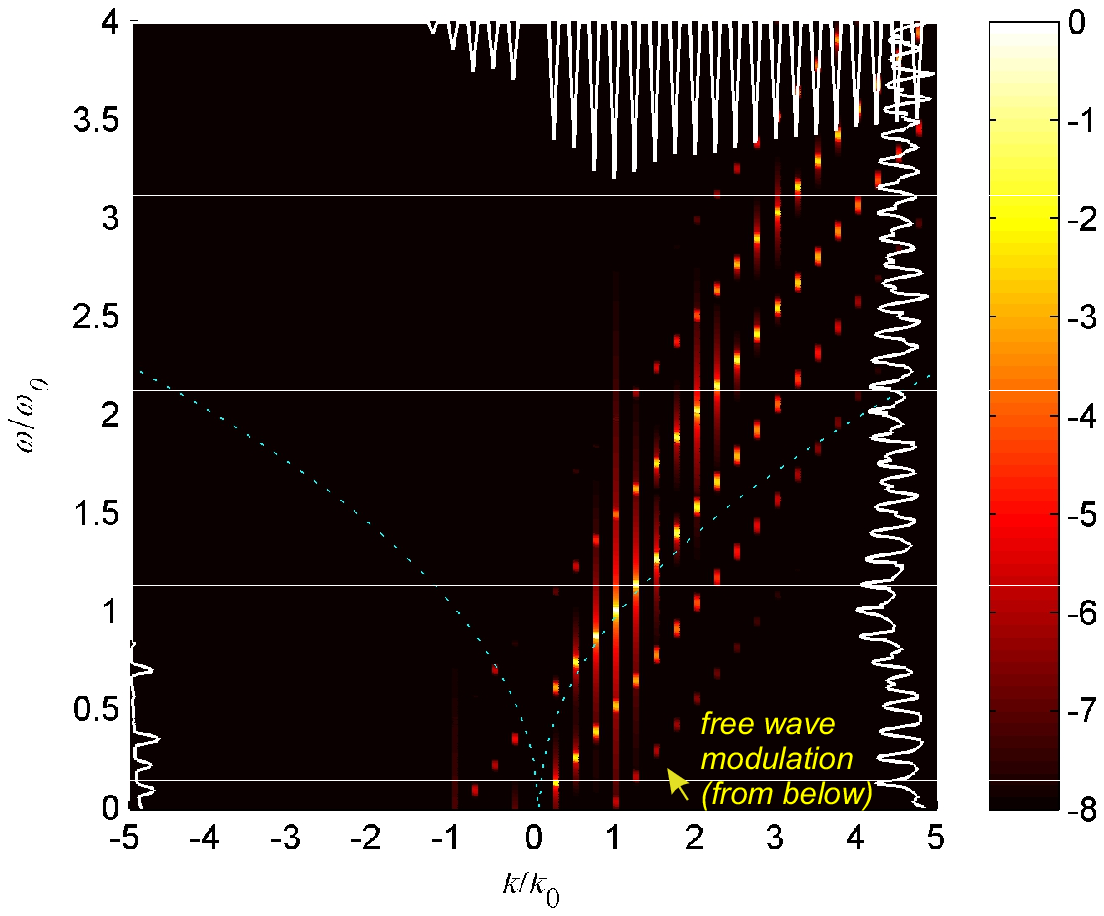}(a)}
\centerline{\includegraphics[width=8cm]{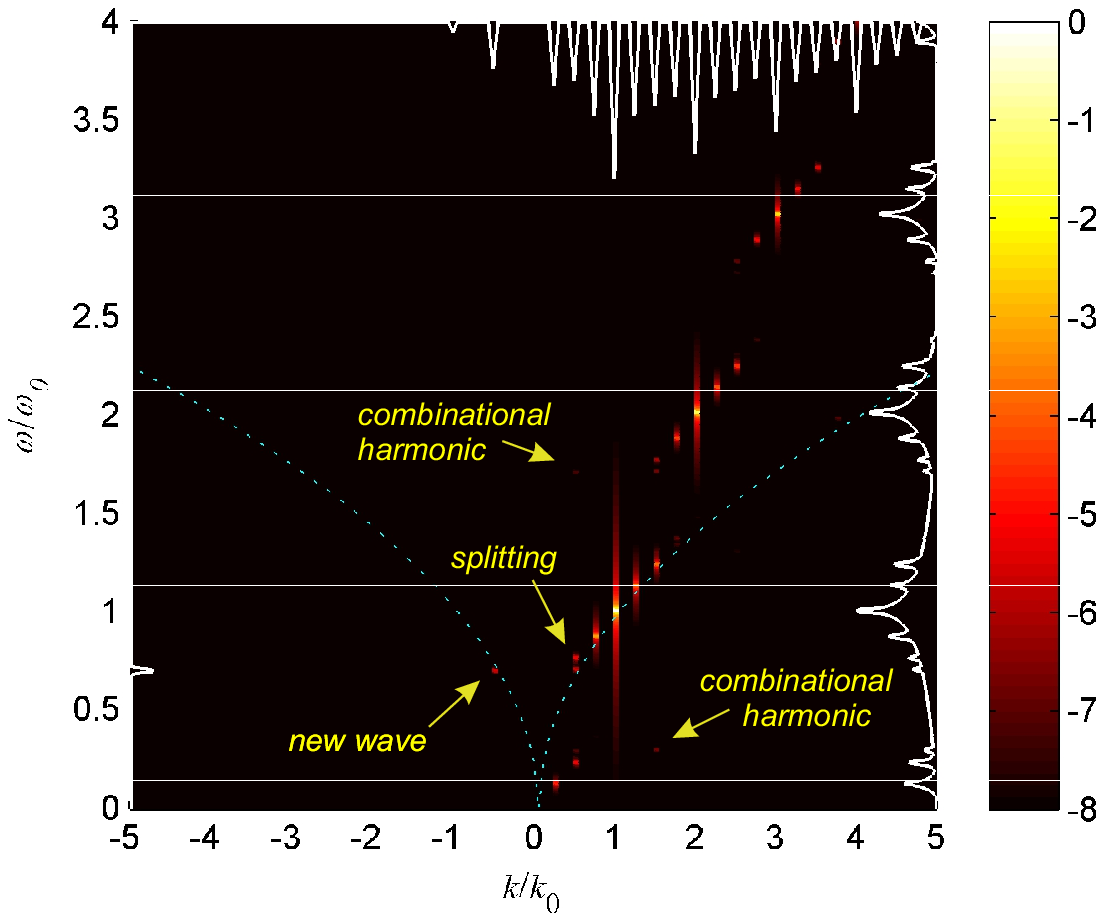}(b)}
\caption{The Fourier transform amplitudes at the moment of focusing ($t \approx 1600 T_0$) (a) and well after ($t \approx 2000 T_0$) (b), the case $\epsilon = 0.11$, $L/\lambda=4$. The notations are similar to Fig.~\ref{fig:XTSp007}. See clip \cite{MovieCase011}.}
\label{fig:XTSp011}
\end{figure}

\begin{figure}
\centerline{\includegraphics[width=8cm]{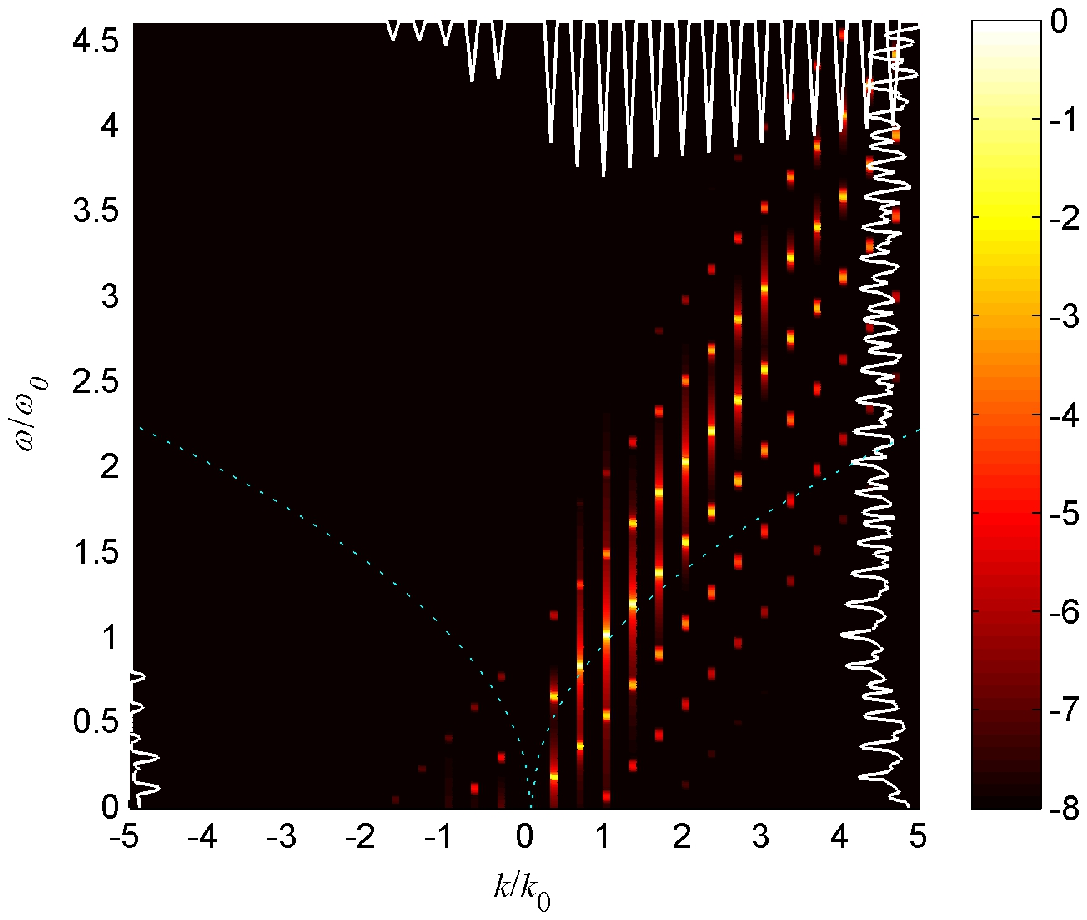}(a)}
\centerline{\includegraphics[width=8cm]{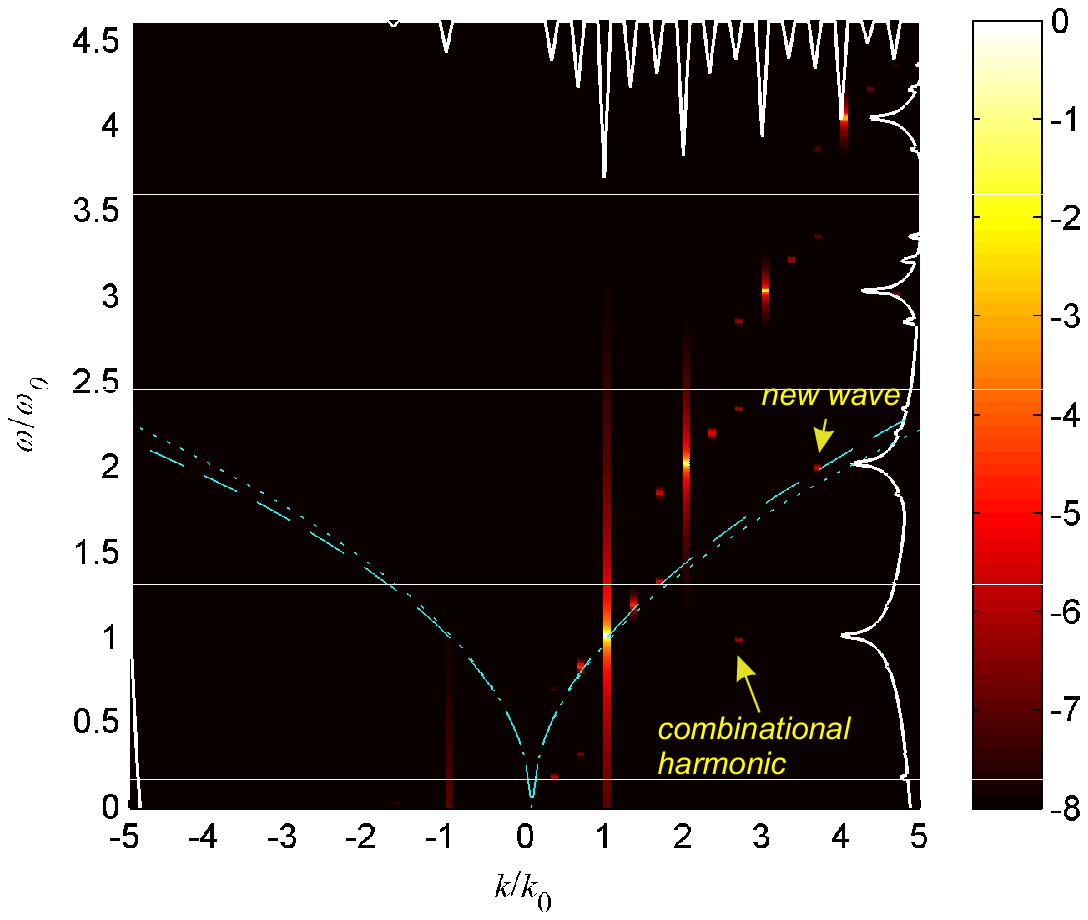}(b)}
\caption{The Fourier transform amplitudes at the moment of focusing ($t \approx 1500 T_0$) (a) and well after ($t \approx 1900 T_0$) (b), the case $\epsilon = 0.146$, $L/\lambda=3$. The cyan broken lines in panel (b) plot the modified dispersion curve. The other notations are similar to Fig.~\ref{fig:XTSp007}, \ref{fig:XTSp011}. See clip \cite{MovieCase0146}.}
\label{fig:XTSp0146}
\end{figure}

Prompted by the graphic example in Fig.~\ref{fig:XTSp007}, one may anticipate that the energy to the opposite waves could also be transferred by the bound modes of the modulated group. This scenario is illustrated in Fig.~\ref{fig:XTSp011} for the other values of the wave steepness and the length of the modulation. Only the focusing stage and the subsequent defocusing are shown in the figure.

At the maximal modulation stage (Fig.~\ref{fig:XTSp011}a) the spots in the domain $k<0$ are represented by two lobes of the modulation.
The first one corresponds to the modulation of the free wave, while the second corresponds to the modulation of the second harmonic. Two local maxima of $S$ are located in the vicinity of the dispersive curve: at $k/k_0=-1/2$ and $k/k_0=-1/4$.  When the modulation relaxes (Fig.~\ref{fig:XTSp011}b), only the wave excited by the second harmonic remains ($k/k_0=-1/2$) with the amplitude about $O(\epsilon^{-6})$ of the amplitude of the carrier wave.
The lengths of the generated waves presented in Fig.~\ref{fig:XTSp007}b,c and Fig.~\ref{fig:XTSp011}b agree with the estimation for a quasi-linear wave group characterized by the dominant wavenumber $k_0$, frequency $\omega_0$ and velocity $\omega_0/(2k_0)$.


It is obvious that interactions with higher-order nonlinear harmonics of the modulated Stokes could generate opposite waves with other lengths. The higher-order harmonics are smaller in amplitude, but at the same time their Fourier spectra are wider for the higher nonlinearity. As a result, the patterns of energy in the $k-\omega$ planes which represent different harmonics seem to be of similar sizes.


\emph{Generation of waves in the same direction.---}
In Fig.~\ref{fig:XTSp0146} we present the case when the induced large-scale displacement (in other words, the zeroth harmonic) generates the linear wave which travels in the same direction as the main train.
%
In the limit of vanishing nonlinearity the zeroth harmonic of the wave group with velocity $\omega_0/(2k_0)$ should cross the linear dispersion curve at $(4k_0, 2\omega_0)$.

In Fig.~\ref{fig:XTSp0146}a the picture of the waves excited due to the modulation  is even more stirring then previously. However, the opposite waves are not generated, only spurious peaks are present in $S_k$ at $k<0$ in Fig.~\ref{fig:XTSp0146}b as they are not accompanied by the peaks in $S_{\omega-}$. At the same time, the local maximum of $S$ near the dispersive curve at $k=4k_0$ is rather well seen in Fig.~\ref{fig:XTSp011}b (actually, with a slightly smaller wavenumber, $k=11/3k_0$), it has the amplitude about $O(\epsilon^{-6})$ of the dominant wave amplitude. In fact, a similar spot is present in Fig.~\ref{fig:XTSp011}b, though with about one order smaller amplitude.

We should emphasize that the quasi-linear waves generated by the  dominant strongly modulated wave train have the frequencies which are noticeably different from the linear solution (this is especially clear in Fig.~\ref{fig:XTSp0146}b).
The intense wave causes the nonlinear frequency upshift for the waves moving in the same direction and the downshift for the opposite waves, $\omega_{nl} \approx \sqrt{gk} + \epsilon^2 \omega_0 k/k_0$ for $|k|>k_0$  (e.g., in \cite{Zakharov1999}). Here $\epsilon$ is the steepness of the initial Stokes wave. The modified dispersive curve $\omega_{nl}(k)$ is shown by the cyan broken line in Fig.~\ref{fig:XTSp0146}b; a good fit with the simulated waves is clear.



We may also note the following peculiarities of the observed $k-\omega$ Fourier transforms.
In Fig.~\ref{fig:XTSp011}a and Fig.~\ref{fig:XTSp0146}a the modulations are so strong that the tails of the modulation lobes in the quarter $k<0$ cross the horizontal axis and continue in the quarter $k>0$ from below. This may lead to the complicated shape of the frequency spectrum $S_{\omega+}$ in the low-frequency range.
A splitting of the peak near $(k_0/2, \omega_0/\sqrt{2})$ may be seen in Fig.~\ref{fig:XTSp011}b, which reveals that some energy from the nonlinear modulation at $k_0/2$ has been transferred to the linear wave with the same wavenumber. This effect is discussed in more detail in \cite{SlunyaevDosaev2017}.
The combinational harmonics caused by the interaction between the main harmonic $(k_0, \omega_p)$ and the new generated waves are visible in Fig.~\ref{fig:XTSp011}b and Fig.~\ref{fig:XTSp0146}b .

\emph{Conclusions.---}
In this work we study the essentially nonlinear wave interactions of strongly modulated waves with the help of the representation in the $k-\omega$ Fourier space. A particular example of potential water waves is simulated numerically with the high accuracy. The observed picture of the frequency-wavenumber Fourier domain is found to be much more complicated than generally assumed even in the utmost simplified problem setup.

We found that the self-modulation of the water wave trains leads to the formation of coherent nonlinear wave groups, which in the Fourier space essentially deviate from the dispersive relation. This excursion from the dispersive curve results in excitation of new waves due to the Cherenkov-type resonance (\ref{CherenkovResonanceGroup}).  Besides, the wave groups are found able to radiate through the phase-locked bound waves, what greatly enriches the possible spectrum of radiated waves and complicates the problem of stability of quasisolitons.
According to the conventional nomenclature, the presented new wave interactions may be referred to the strongly non-resonant wave interactions or to a new type of a wave-group resonance or of nonlinear Cherenkov radiation.
The general condition on the resonance between waves and the coherent wave group with a velocity $V$, dominant wavenumber $k_0$ and frequency $\omega_p$ is
\begin{align} \label{NewResonantCondition}
V = (\omega - m \omega_p) / (k- m k_0) ,
\end{align}
where $m=0,1,2,...$ numerates the branch of the resonance. The classic Cherenkov radiation (\ref{CherenkovResonanceGroup}) occurs for $m=1$.

The revealed effect of generation of new waves by the modulated train seems to be not captured by the Zakharov equation for the effective Hamiltonian \cite{Zakharov1999} as in the presented simulations the difference between free (true, natural) and bound (phase-locked) modes becomes indistinct. The employed inherently linear Fourier analysis cannot help to differentiate them.

In the considered example of surface water waves the amplitudes of the radiated waves are several orders of magnitude smaller than of the dominant wave, thus in the general case these wave interactions may be concealed by other stronger ones. It is clear that the imposed discreteness of the wavenumbers impedes the interactions; they would have been more efficient if this restriction was absent. The realizability of the similar dynamics caused by higher-order nonlinearities is obvious.
It is interesting to note that the patterns of energy which represent different bound wave harmonics in the $k-\omega$ plane seem to be of similar sizes (see the figures), what can be explained by the nonlinear mechanism of their appearance. Though the higher-order nonlinear harmonics are characterized by smaller amplitudes ($\sim \epsilon^n$, where $n=1$ corresponds to the dominant wave), at the same time their Fourier transforms are relatively wider. Therefore the comparative significance of the higher order resonances described by the condition (\ref{NewResonantCondition}) depends mainly on the mutual location in the $k - \omega$ plane of the nonlinear group harmonics and of the dispersion curves for linear waves.

We should mention that the considered situation of collinear water waves is well-known to be degenerative in comparison with the directional wave situation, as many resonant interactions become forbidden (e.g. \cite{Zakharov1999}). In the general statement of directional waves the dynamics will be obviously much more complicated. Meanwhile the reasons why the observed interactions may be canceled, are not seen a priori.
The discovered effects are of the universal nature and should be observable in many other nonlinear media with dispersion.

The support from RSF Grant No. 16-17-00041 is acknowledged. The author is grateful to E.N.~Pelinovsky, V.I.~Shrira, E.A.~Kuznetsov, A.I.~Dyachenko, D.I.~Kachulin for stimulative discussions, and to the anonymous referee for noticing the alternative form of interpretation of (\ref{CherenkovResonanceGroup}) as the equality of the Doppler-shifter frequencies.

\end{document}